\documentclass[pra,twocolumn,showpacs,amsmath,amssymb,superscriptaddress]{revtex4}
\usepackage{graphicx}
\usepackage{natbib}
\usepackage{latexsym}

\begin{document}

\title{Modifying molecule-surface scattering by ultrashort laser pulses}

\author{Yuri Khodorkovsky}
\affiliation{Department of Chemical Physics, The Weizmann Institute of Science, Rehovot 76100, Israel}

\author{J. R. Manson}
\affiliation{Department of Physics and Astronomy, Clemson University, Clemson, SC 29634, USA}

\author{Ilya Sh.\ Averbukh}
\affiliation{Department of Chemical Physics, The Weizmann Institute of Science, Rehovot 76100, Israel}

\begin{abstract}
In recent years it became possible to align molecules in free space using ultrashort laser pulses. Here we explore two schemes for controlling molecule-surface scattering process, which are based on the laser-induced molecular alignment. In the first scheme, a single ultrashort nonresonant laser pulse is applied to a molecular beam  hitting the surface. This pulse modifies the angular distribution of the incident molecules, and causes the scattered molecules to rotate with a preferred sense of rotation (clockwise or counter-clockwise). In the second scheme, two properly delayed laser pulses are applied to a molecular beam composed of two chemically close molecular species (isotopes, or nuclear spin isomers). As the result of the double pulse excitation, these species are selectively scattered to different angles after the collision with the surface. These effects may provide new means for the analysis and separation of molecular mixtures.
\end{abstract}
\pacs{37.10.Vz,34.35.+a,34.50.Rk,68.49.Df}
\maketitle

\section{Introduction}

Laser control of molecular rotation, alignment and orientation has received significant attention in recent years (for a review, see e.\ g.\ \cite{Stapelfeldt03,Tamar06}). Interest in the field has increased, mainly due to the improved capabilities to manipulate the characteristics of the laser pulses (such as time duration and temporal shape), which in turn leads to potential applications offered by controlling the angular distribution of molecules. Since the typical rotational time scale is 'long' (${\sim}10 \,\mathrm{ps}$) compared to the typical short pulse duration (${\sim}50 \,\mathrm{fs}$), effective rotational control and manipulation are in reach.
During the last decade, coherent rotational dynamics of pulse-excited molecules was studied \cite{Ortigoso,Rosca-Pruna}, and multiple pulse sequences giving rise to the enhanced alignment were suggested \cite{Averbukh,Leibscher,Renard}, and realized experimentally \cite{Bisgaard1,Lee,Bisgaard2,Pinkham}. Further manipulations, such as optical molecular centrifuge and alignment-dependent strong field ionization of molecules, were demonstrated \cite{Karczmarek,Litvinyuk}. Selective rotational excitation in bimolecular mixtures was suggested and demonstrated in the mixtures of  molecular isotopes \cite{Fleischer06} and molecular spin isomers \cite{Faucher,Fleischer07}.
These new methods for manipulation of molecular rotation can also be used to modify the motion of molecules in inhomogeneous fields, such as focused laser beams \cite{Purcell09,Gershnabel10,frisbee}, or static electric \cite{electric} and magnetic \cite{magnetic} fields.

A molecule near a solid surface can be also considered as a particle in a complicated inhomogeneous field.
Although the potential energy of a molecule near a solid surface is quite complicated in all its details, there are cases when simple potential models can be used. We treat below the molecule as a rigid rotor \cite{Manson06}, while the surface is considered flat and is described by a hard cube model \cite{Logan66,Doll73}. This model was used and gave qualitatively correct results for $\text{N}_2$ or $\text{NO}$ molecules incident with thermal velocities on close-packed surfaces, such as $\text{Ag}(111)$ (see reference \cite{Spruit89} and references therein).

Modification of the molecule-surface scattering and molecule-surface reactions by external fields of different nature is a long-standing research problem. In particular, the effect of molecular orientation by a static electric hexapole field on the scattering process was investigated in detail \cite{Kleyn}. Laser control of the  gas-surface scattering was achieved using multiphoton ionization of the impinging molecules by long laser pulses of variable polarization \cite{Jacobs}, and possibility of controlling  molecular adsorption on solid surfaces using ultrashort laser pulses was discussed \cite{Seideman_surf}.

In this paper, we investigate the prospects of modifying and controlling the process of molecular scattering  from solid surfaces by using ultrashort laser pulses that align molecules before they hit the surface.

The paper is organized as follows. The molecule-surface collision model is presented in Sec.\ II. Next, in Sec.\ III, the model for the interaction of a molecule with an ultrashort laser pulse is briefly explained. Then, we suggest two schemes for laser control of the molecular scattering process. In the first one, explained in Sec.\ IV, a single laser pulse is applied to a molecular beam in order to cause the  molecules to rotate with a chosen sense of rotation (clockwise or counterclockwise) after the scattering from the surface. In the second suggested scheme, explained in Sec.\ V, two properly delayed laser pulses are applied to a molecular beam composed  of two chemically close molecular species (isotopes, or nuclear spin isomers). As a result of the double pulse excitation, the subspecular scattering angles become enriched in one of these species after the scattering from the surface. In Sec.\ VI we summarize and conclude.

\section{Molecule-surface scattering model}




In this paper, we use a model in which molecule is treated as a rigid dumbbell \cite{Doll73}. This dumbbell collides with a flat frictionless hard cube, that represents one of the  surface atoms \cite{Logan66}. We assume that the cube has some velocity that is distributed according to the surface temperature. This hard cube model provides a simple way of adding surface phonons to the molecule-surface collision process. For the sake of simplicity, we also assume that the cube is much heavier than the molecule, so that its velocity does not change as a result of the collision. In this case, by moving to the frame attached to the cube, one reduces the problem to the molecular collision with a motionless hard wall. In the moving coordinate system, the molecular total energy (translational+rotational) is conserved, but it can be redistributed between these two parts as a result of the collision.

What about the translational linear momentum of the molecule in the moving coordinate system? The component perpendicular to the surface is not conserved, because the surface exerts forces on the molecule in this direction during the collision. On the other hand, there are no forces applied in the direction {\it parallel} to the frictionless surface, and therefore the linear momentum parallel to the surface is conserved. This is the reason why we can simplify the problem and consider a colliding molecule that has only a translational velocity component perpendicular to the surface. Notice that this cannot be done if the surface is corrugated.

Using energy and angular momentum conservation laws (as explained below), we find analytic expressions for the translational and the rotational velocities of the dumbbell molecule after the collision. These velocities depend on the velocities before the collision, and on the angle between the dumbbell and the surface of the cube at the moment of collision. Finally, we transform the velocities back to the laboratory coordinate frame.

In the next subsections, we treat a simple case of a homonuclear molecule rotating in a plane and colliding with a heavy hard cube. Next, we extend the treatment to a 3-dimensional rotation of a heteronuclear molecule.  In the last subsection, we consider the effects of the surface cube vibration on the collision.

\subsection{The two-dimensional collision of a homonuclear diatomic molecule}

\begin{figure}
\centering
\includegraphics[width=0.25\textwidth]{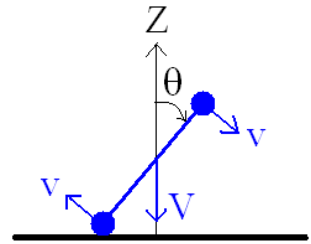}
\caption{A homonuclear diatomic molecule hitting a flat hard surface. This is a two-dimensional model, where the molecule rotates only in the plane of the figure. The translational velocity of the center of mass is $V$, the rotational velocity of each one of the atoms, in the coordinate system moving with the center of mass, is $v$, and the angle between the surface normal (the $Z$-axis) and the molecular axis is $\theta$.}
\label{dumbbell}
\end{figure}

We treat a homonuclear diatomic molecule as a massless stick of length $r_e$, with two atoms, each of mass $m$, attached to its ends.
To describe the molecular motion, we define the $Z$-axis perpendicular to the surface, see Fig.\ \ref{dumbbell}. The angle between the molecular axis and the $Z$-axis is $\theta$, and it belongs to the range of $[0,2\pi]$. The translational velocity of the center of mass is denoted by $V$. The linear rotational velocity is denoted by $v$, and is equal to $r_e\omega/2$, where $\omega$ is the angular velocity of the molecular rotation.
The velocities $V$ and $v$ are defined in the coordinate system moving with the velocity of the surface cube.

{\bf Energy conservation.} The total energy conservation for the molecule is:
\begin{equation}
\label{cons_energy-2}
\frac12 (2m)V_i^2+\frac12 I\omega_i^2=\frac12 (2m)V_f^2+\frac12 I\omega_f^2~,
\end{equation}
where $I=mr_e^2/2$ is the moment of inertia, and the subscripts $i$ and $f$ denote the velocities before and after the collision, respectively. Simplifying the above expression we obtain:
\begin{equation}
\label{cons_energy-1}
V_i^2+v_i^2=V_f^2+v_f^2~,
\end{equation}
or:
\begin{equation}
\label{cons_energy}
(V_i-V_f)(V_i+V_f)=(v_f-v_i)(v_f+v_i)~.
\end{equation}

{\bf Angular momentum conservation.} The angular momentum of this system of two particles depends on the choice of the coordinate system. We choose it such that at the moment of collision the origin is at the position of the colliding atom. Thus, the other atom is the only one contributing to the angular momentum of the system. Moreover, the torque exerted by the wall is zero for this choice of the coordinate system. Therefore, {\it for this choice of the coordinate system}, the angular momentum is conserved during the collision. Adding the velocities $V$ and $v$, as defined in Fig.\ \ref{dumbbell}, and assuming $\theta$ being between $0$ and $\pi/2$, we equate the magnitude of the angular momentum before and after the collision:
\begin{equation}
\label{cons_ang_momentum-1}
mr_e(v_i-V_i\sin{\theta})=mr_e(v_f-V_f\sin{\theta})~,
\end{equation}
or:
\begin{equation}
\label{cons_ang_momentum}
(V_i-V_f)\sin{\theta}=v_i-v_f~.
\end{equation}
Dividing (\ref{cons_energy}) by (\ref{cons_ang_momentum}) and using the result together with Eq.\ (\ref{cons_ang_momentum}), we obtain the expressions for $V_f$ and $v_f$:
\begin{equation}
V_f=\frac{-V_i\cos^2{\theta}-2v_i\sin{\theta}}{1+\sin^2{\theta}}~,\nonumber
\end{equation}
\begin{equation}
v_f=\frac{v_i\cos^2{\theta}-2V_i\sin{\theta}}{1+\sin^2{\theta}}~.
\end{equation}
These equations should give the same results for $\theta\to\theta+\pi$, because of the symmetry of the molecule. Because the sine function changes sign under this transformation, the correct equations for $\theta$ in the first or the third quadrant, are:
\begin{equation}
V_f=\frac{-V_i\cos^2{\theta}-2v_i|\sin{\theta}|}{1+\sin^2{\theta}}~,\nonumber
\end{equation}
\begin{eqnarray}
\label{Vvf_1}
v_f=\frac{v_i\cos^2{\theta}-2V_i|\sin{\theta}|}{1+\sin^2{\theta}}~; \\
\mathrm{for}\,\, 0\le\theta<\pi/2\,\, \mathrm{and}\,\, \pi\le\theta<3\pi/2 \nonumber
\end{eqnarray}
For the angle $\theta$ in the second or the fourth quadrant, the transformation $\theta\to\pi-\theta$ should be done. This is equivalent to changing $v_i$ and $v_f$ to $-v_i$ and $-v_f$. Finally, we obtain:
\begin{equation}
V_f=\frac{-V_i\cos^2{\theta}+2v_i|\sin{\theta}|}{1+\sin^2{\theta}}~,\nonumber
\end{equation}
\begin{eqnarray}
\label{Vvf_2}
v_f=\frac{v_i\cos^2{\theta}+2V_i|\sin{\theta}|}{1+\sin^2{\theta}}~; \\
\mathrm{for}\,\, \pi/2<\theta\le\pi\,\, \mathrm{and}\,\, 3\pi/2<\theta\le 2\pi \nonumber
\end{eqnarray}
Notice that $V_i$ is always negative, and $v_f$ is positive/negative for a clockwise/counterclockwise rotation, respectively.

The angles $\theta{=}\pi/2$ or $3\pi/2$ are unique and are excluded from the equations above. The reason is that the treatment above is incorrect for this angle of incidence, because both atoms hit the surface at the same time, and the law of conservation of angular momentum, as applied above, cannot be applied in this specific case. However, these angles of incidence can be treated as limiting cases of a double collision, and treating them provides no particular problem, as explained below.

{\bf Time evolution.} The above analytical expressions for translational and rotational velocities describe a {\it single} collision of the molecule with the surface. However, {\it additional collisions} may occur as well. This scenario is most evident for molecules hitting the surface at the angle close to $\pi/2$. In the case of the additional collision, the equations (\ref{Vvf_1}) and (\ref{Vvf_2}) should be applied again.

Instead of finding complicated conditions for multiple collisions, we turn to a simple numerical simulation. The equations of motion for the molecule, starting at some $Z_0$ and $\theta_0$ with velocities $V_i$ and $v_i$, before the first collision, are simple:
\begin{equation}
\label{Zt}
Z(t)=Z_0+V_it~,
\end{equation}
and
\begin{equation}
\label{th_t}
\theta(t)=\theta_0+\omega_0t=\theta_0+\frac{2}{r_e}v_it,\mod{(2\pi)}
\end{equation}
The collision occurs when the distance between the face of the cube and the center of mass of the molecule is equal to $0.5 r_e|\cos{\theta (t)}|$.
At this time moment, the velocities are transformed according to (\ref{Vvf_1}) or (\ref{Vvf_2}), and the evolution continues according to (\ref{Zt}) and (\ref{th_t}), using the new values of $Z_0$, $V_i$, $\theta_0$ and $v_i$. We assume here that the collision lasts for a very short period of time, such that during the collision the values of $Z$ and $\theta$ are practically constant.

\subsection{Extension to three-dimensional collision of a heteronuclear diatomic molecule}

\begin{figure}
\centering
\includegraphics[width=0.5\textwidth]{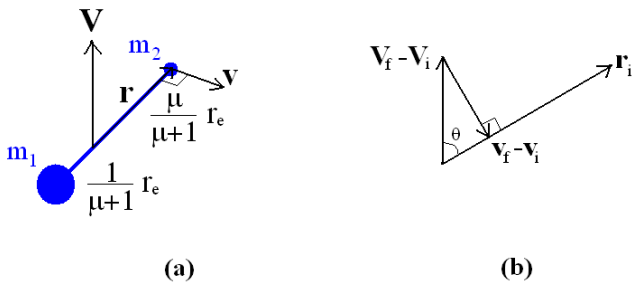}
\caption{(a) A heteronuclear diatomic molecule composed of atoms of masses $m_1$ and $m_2$ (where $m_1>m_2$) connected by a massless rod of length $r_e$. The translational velocity of the molecular center of mass is $\mathbf{V}$, the molecular orientation in space is represented by the vector $\mathbf{r}$, and the corresponding rotational velocity of the light atom is $\mathbf{v}$. (b) A vector diagram corresponding to Eq.\ (\ref{cons_ang_momentum_3Da}), where molecular orientation vector at the moment of collision is denoted by $\mathbf{r}_i$.}
\label{dumbbell_3D}
\end{figure}

We describe the heteronuclear molecule as composed of two different atoms of masses $m_1$ and $m_2$, where $m_1>m_2$. The atoms are connected by a massless rod of length $r_e$, as shown in Fig.\ \ref{dumbbell_3D}(a). The center of mass of the molecule is closer to the heavy atom, and is at the distance of $r_e/(\mu+1)$ from it, according to the definition of the center of mass. Here $\mu$ was defined as $m_1/m_2$. The distance between the center of mass and the light atom is, accordingly, $r_e\mu/(\mu+1)$.

The molecule is characterized by the center of mass coordinate $Z$, and the direction of the molecular axis, described by a vector $\mathbf{r}=\frac{\mu}{\mu+1}r_e(x,y,z)$, where $(x,y,z)$ is a unit vector. This vector points from the molecular center of mass to the light atom of mass $m_2$, see Fig.\ \ref{dumbbell_3D}(a). The corresponding velocity is given by $\mathbf{v}=d\mathbf{r}/dt$. Therefore, the linear velocity (in the center of mass coordinate system) of the light atom is $\mathbf{v}$, and that of the heavy atom is $-\mathbf{v}/\mu$.

After a derivation, similar to the one for the 2-D rotation presented above, we obtain expressions connecting velocities before the collision, $\mathbf{V}_i$ and $\mathbf{v}_i$, to the velocities after the collision, $\mathbf{V}_f$ and $\mathbf{v}_f$. The molecule, at the moment of collision, is oriented at $\mathbf{r}_i$, see Fig.\ \ref{dumbbell_3D}(b). The details of the derivation  are given in the Appendix at the end of the paper.

For a free-rotating heteronuclear molecule, the end of the vector $\mathbf{r}$ traces a circle with a radius equal to $\frac{\mu}{\mu+1}r_e$. The orientation of the circle in space is defined by the initial orientation $\mathbf{r_0}$ and the initial velocity $\mathbf{v_0}$. The orientation of the molecule at time $t$ is thus given by:
\begin{equation}
\label{r_of_t}
\mathbf{r}(t)=\mathbf{r_0}\cos{(v_0t)}+\frac{\mu}{\mu+1}r_e\frac{\mathbf{v_0}}{v_0}\sin{(v_0t)}~,
\end{equation}
where $v_0=|\mathbf{v_0}|$. Taking the derivative, the velocity is:
\begin{equation}
\label{v_of_t}
\mathbf{v}(t)=-v_0\mathbf{r_0}\sin{(v_0t)}+\frac{\mu}{\mu+1}r_e\mathbf{v_0}\cos{(v_0t)}~.
\end{equation}

Finally, we emphasize that the molecular collision with the surface of the cube occurs when the molecule is oriented with $z_i>0$ and the distance between the face of the cube and the center of mass of the molecule is equal to $r_e|\cos{\theta}|/(\mu+1)$, {\bf or} when the molecule is oriented with $z_i<0$ and the distance between the face of the cube and the center of mass of the molecule is equal to $r_e|\cos{\theta}|\,\mu/(\mu+1)$.

\subsection{Vibration of surface atoms - simple inclusion of phonons}

The hard cube model \cite{Goodman65,Logan66,Doll73} provides a simple way for including surface atom vibration into the collision process.
The collision of a molecule with a hard wall of infinite mass is replaced by a collision with a hard cube of a finite, but a large mass $M$ moving with velocity $U$. For simplicity, we assume that the cube oscillates in a hard box of a finite size, which is a free parameter of the model. In the model, the cube oscillates in the direction perpendicular to the surface plane. The reason is that for a flat and frictionless cube only the vertical velocity component can transfer energy to the impinging molecule. The cube moves with a constant speed, while the velocity reverses its direction at the ends of the hard box. The one dimensional velocity of the hard cube is random and is distributed according to a Boltzmann distribution:
\begin{equation}
\label{surface_U_dist}
f(U)=\sqrt{\frac{M}{2\pi k_BT_{\text{surf}}}}\exp{\left(\frac{MU^2}{2k_BT_{\text{surf}}}\right)}~,
\end{equation}
where $T_{\text{surf}}$ is the temperature of the solid surface.
When treating the collision process, we assume that the mass of the surface atom is much larger than the mass of the molecule, i.\ e.\ $M\gg m_1+m_2$, so that the collision does not change significantly the velocity of the hard cube. Only the velocities of the molecule change, and its total energy in the laboratory coordinate frame may increase or decrease.

The previous treatment of the collision in the coordinate frame moving with the cube is now easily incorporated into the model. The collision condition (for a homonuclear molecule) can be expressed as
\begin{equation}
\label{condition_phonons}
Z(t)-Z_M(t)=\frac12 r_e |\cos{\theta(t)}|~,
\end{equation}
where $Z_M(t)$ is the time-dependent position of the surface of the hard cube. According to the assumptions above, $Z_M(t)$ is a simple ``zigzag'' function (triangle wave) with an amplitude that is a free parameter, which we take equal to the molecular bond length $r_e$, and a frequency determined by the cube speed $|U|$. The translational velocity of the molecule is transformed before the collision according to $V_i\to V_i-U_{col}$, where $U_{col}$ is the velocity of the hard cube at the moment of collision. This is a usual Galilean transformation to the coordinate frame moving with the hard cube. Similarly, after the collision the translational velocity of the molecule is transformed back by $V_f\to V_f+U_{col}$. The rotational velocity remains unchanged under this Galilean transformation.

\section{Interaction of the molecule with an ultrashort laser pulse}

Here we briefly summarize the results of the classical model describing the interaction of the diatomic rigid molecule with a nonresonant ultrashort laser pulse, in the impulsive approximation. A more detailed description may be found in \cite{Khodorkovsky}.

The potential energy of the laser pulse interacting with the induced molecular dipole is given by:
\begin{equation}
\label{V}
V(\theta,\varphi,t)=-\frac{1}{4}\mathcal{E}^2(t)\left(\Delta\alpha\cos^2{\beta}+\alpha_{\perp}\right)~,
\end{equation}
where $\Delta\alpha=\alpha_{\parallel}-\alpha_{\perp}$ is the difference between the polarizability along the molecular axis and the one perpendicular to it, $\mathcal{E}(t)$ is the envelope of the electric field of the {\it linearly polarized} laser pulse, and $\beta=\beta(\theta,\varphi)$ is the angle between the molecular axis and the direction of polarization of the pulse. Here $\theta,\varphi$ are the polar and the azimuthal angles characterizing the orientation of the molecular axis, respectively.  We assume that the pulse duration is very short compared to the rotational period, so that the pulse can be described in the impulsive ($\delta$-kick) approximation. We define the dimensionless interaction strength $P$, which characterizes the pulse, as
\begin{equation}
\label{P}
P=\frac{\Delta\alpha}{4\hbar}\int_{-\infty}^{\infty}\mathcal{E}^2(t)dt~.
\end{equation}

We consider the action of a pulse linearly polarized along some arbitrary unit vector $\mathbf{p}$, and determine the vector of the resulting velocity change $\boldsymbol{\Delta}\mathbf{v}$ for a molecule oriented along some direction $\mathbf{r_0}$. The norm $|\boldsymbol{\Delta}\mathbf{v}|$ can be found by integrating Newton's equations of motion for a pulse polarized along the $z$-axis. It is equal to $\frac{\hbar}{I}|P\sin{2\beta_0}|$, where $\beta_0$ is the angle between the polarization direction of the pulse $\mathbf{p}$ and the orientation direction of the molecule $\mathbf{r_0}$, and $I$ is the moment of inertia of the molecule. Notice that $\boldsymbol{\Delta}\mathbf{v}$ is always perpendicular to $\mathbf{r_0}$. Also, $\boldsymbol{\Delta}\mathbf{v}$ is directed parallel or antiparallel to the vector component of $\mathbf{p}$ perpendicular to $\mathbf{r_0}$, which is equal to $\mathbf{p}-\frac{\mu+1}{\mu r_e}\cos{\beta_0}\mathbf{r_0}$. As a result, we arrive at:
\begin{equation}
\label{delta_v_vec}
\boldsymbol{\Delta}\mathbf{v}=\frac{2\hbar P}{I}\cos{\beta_0}\left(\frac{\mu}{\mu+1}r_e\mathbf{p}-\cos{\beta_0}\mathbf{r_0}\right)~.
\end{equation}

\section{``Molecular propeller'' induced by laser alignment and collision with the surface}

In this section we explore a way of inducing unidirectional molecular rotation by a single laser pulse and a single surface scattering event. The idea is inspired by a recent scheme that was proposed in \cite{Fleischer09,York09} and realized experimentally in \cite{Kitano09}. In these papers, two time-delayed and cross-polarized laser pulses were used in order to induce molecular rotation of a preferred sense. Here we achieve a similar goal by replacing the second laser pulse by the process of molecular scattering from a solid surface.

Before presenting our new scheme, we summarize shortly the pure optical one described in \cite{Fleischer09,York09,Khodorkovsky,Kitano09}. We start with a gas of diatomic molecules in free space. The first ultrashort laser pulse, linearly polarized along the $z$-axis, induces coherent molecular rotation that continues after the end of the pulse.  The molecules rotate under field-free conditions until they reach an aligned state, in which the molecular axis with the highest polarizability is confined to a narrow cone around the polarization direction of the first pulse. The second short laser pulse is applied at the moment of the best alignment, and at angle with respect to the first pulse. As a result, the aligned molecular ensemble experiences a torque causing molecular rotation in the plane defined by the two polarization vectors. The rotational velocity delivered to a linear molecule is maximal when the laser pulse is polarized at 45 degrees with respect to the molecular axis of the highest polarizability, as can be seen from Eq.\ (\ref{delta_v_vec}). This defines the optimal angle between the laser pulses. The direction of the excited rotation (clockwise or counter-clockwise) is determined by the sign of the relative angle ($\pm45$ degrees) between the first and the second pulse in the polarization plane. This double pulse scheme was termed ``molecular propeller'', as it resembles the action needed to ignite a rotation of a plane propeller.

In the current scheme, we start from a monoenergetic molecular beam of diatomic molecules flying towards a flat surface at the incidence angle of $45^{\circ}$. Before hitting the surface, the molecules are aligned by a laser pulse polarized at  $+45^{\circ}$ to the surface. When colliding with the surface, the aligned molecules receive a ``kick'' from it, and scatter with rotation in a specific direction. Changing the polarization angle of the laser with respect to the surface to  $-45^{\circ}$, the sense of rotation of the scattered molecules can be inverted.
\begin{figure}
\centering
\includegraphics[width=0.5\textwidth]{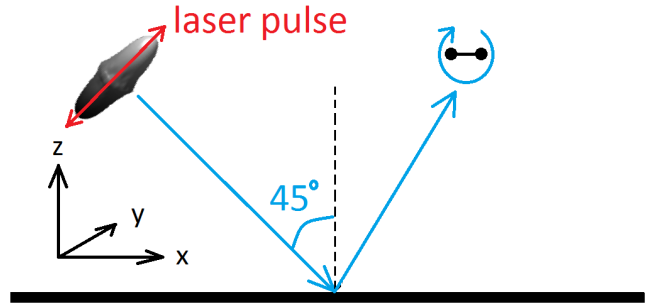}
\caption{(Color online) A molecular beam impinges on a hard-wall surface at  angle of $45^{\circ}$. Before hitting the surface, the molecules are ``kicked'' by an ultrashort laser pulse polarized as shown by the double-headed (red) arrow. This pulse generates the {\it time-averaged} angular distribution of the molecular orientation in the form of the ``cigar'' as shown. These aligned molecules are preferentially rotating clockwise after hitting the surface.}
\label{propeller_scheme}
\end{figure}

In order to analyze properly the scattering of a molecular beam, we need to account for the spatial spread of the molecules inside the beam. For typical experimental conditions, this spread is of the order of  $1\,\text{mm}$ in the direction perpendicular to the direction of the molecular beam. This spread is determined by the diameter of the collimating aperture in the experimental system \cite{Pullman90}. A similar spread can be observed in the direction parallel to the direction of the beam propagation, if one uses the pulsed molecular beam technique \cite{Barry86}. We ignore in our calculation the spread in the perpendicular direction and concentrate on the parallel one.

\begin{figure}
\centering
\includegraphics[width=0.5\textwidth]{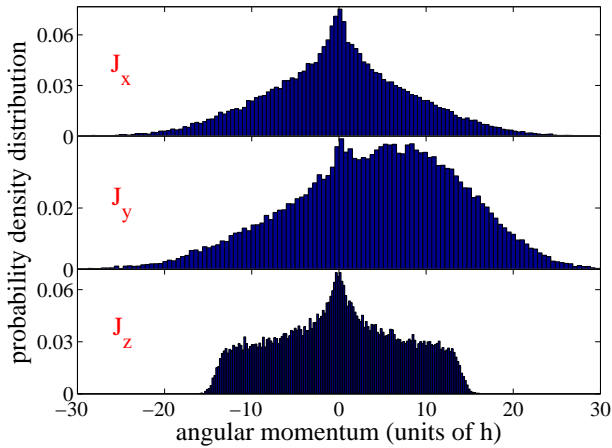}
\caption{The distribution of the angular momentum components $J_x$, $J_y$ and $J_z$ of the molecules scattered from a hard-cube surface are plotted in the three panels. These molecules are manipulated with an ultrashort laser pulse before hitting the surface, as shown in Fig.\ \ref{propeller_scheme}. As expected, the distribution of $J_y$ is asymmetric, and there are more molecules with a positive $J_y$. The average value of $J_y$ is $\langle J_y\rangle=4.2\hbar$. The calculation is done with the hard-cube model for nitrogen molecules (atomic mass of $28\,\text{a.u.}$) incident with translational velocity of $350\,\frac{\text{m}}{\text{sec}}$ and rotational temperature of $1\,\text{K}$. The laser pulse strength is $P=10$, and the polarization angle is $\gamma=45^{\circ}$. The surface is composed of silver atoms at room temperature ($M=108\,\text{a.u.}$ and $T_{\text{surf}}=300\,\text{K}$).}
\label{unidir_Jxyz}
\end{figure}

After the molecules are ``kicked'' by the laser pulse, they start rotating in a concerted way, while continuing to approach the surface. The angular distribution of the ``kicked'' molecules, after averaging over a long time period, is elongated along the pulse polarization direction, as was shown in \cite{Khodorkovsky}, and as is depicted schematically in Fig.\ \ref{propeller_scheme} by the ``cigar''-shaped distribution. The result of this time averaging is the same as of the averaging along the distance parallel to the direction of molecular propagation, because time is related to distance by $Z=V_0t/\sqrt{2}$ for all the molecules. From the above we conclude that {\it on average} the molecules approaching the surface are aligned along the polarization direction of the ultrashort laser pulse. We stress, that this is true classically, as well as quantum mechanically. If the molecules are aligned at the angle of $45$ degrees with respect to the surface, they have a preferred orientation while colliding with the  surface. These molecules  receive a ``kick'' from the surface, which  leads to the preferred sense of rotation, for example, a clockwise rotation, as in Fig.\ \ref{propeller_scheme}. If we plot the distribution of the components of angular momentum of the scattered molecules, we expect to see an asymmetry that is correlated with the polarization direction of the exciting pulse.

In the following, we apply the above scheme to a molecular beam of $\text{N}_2$ molecules hitting a flat hard-cube surface with an angle of incidence of $45^{\circ}$. The molecules move with the initial velocity of $V_0=350\,\frac{\text{m}}{\text{sec}}$, and with the rotational temperature of $1\,\text{K}$, typical for molecular beam experiments. They receive a kick with a strength of $P=10$ from an ultrashort laser pulse, before hitting the surface. We choose the surface to be $\text{Ag}(111)$, with the appropriate mass of the representative hard cube, and the surface temperature is taken to be $300\,\text{K}$.

We choose the polarization direction of the exciting pulse to be $\mathbf{p}=(1,0,1)/\sqrt{2}$, that is, at $45^{\circ}$ to the surface in the $xz$-plane, as shown in Fig.\ \ref{propeller_scheme}. Fig.\ \ref{unidir_Jxyz} shows the distribution functions of the angular momentum components $J_x$, $J_y$ and $J_z$ after the scattering. We see that the distributions of $J_x$ and $J_z$ are symmetric around zero, as expected from the symmetry of the pulse. However, most of the molecules are scattered with positive $J_y$, and its average value is $\langle J_y\rangle=4.2\hbar$. This average is close to the value of of $5.6\hbar$, that may be estimated by using Eq.\ (\ref{Vvf_2}) for $v_f$, and the appropriate molecular constants $m$ and $r_e$. This estimation considers a representative nonrotating molecule oriented at $\theta=\pi/4$ and impinging on the surface with the incident velocity of $V_i=350\, \text{m}/\text{sec}$.

Switching the polarization direction to $\mathbf{p}=(-1,0,1)/\sqrt{2}$  inverts the $J_y$ distribution of the scattered molecules, and makes it peaking at negative values of $J_y$.

\begin{figure}
\centering
\includegraphics[width=0.5\textwidth]{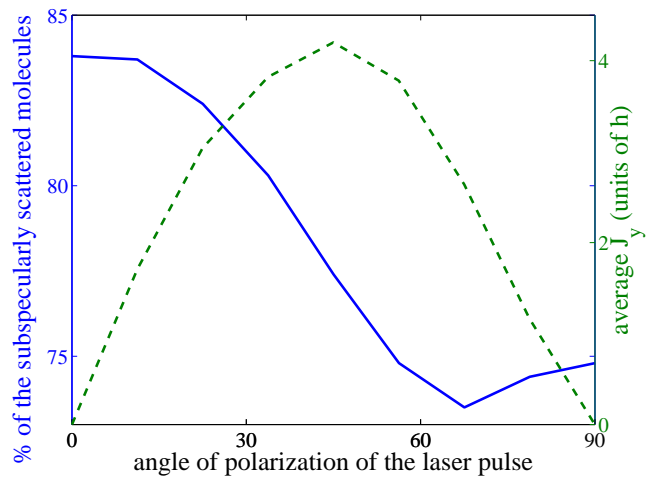}
\caption{(Color online) The percentage of the laser-excited molecules  scattered to the subspecular angles as a function of the polarization angle of the pulse $\gamma$ (solid blue line). The dashed green line represents the average induced angular momentum component $\langle J_y\rangle$ as a function of $\gamma$. The parameters  are the same  as in Fig.\ \ref{unidir_Jxyz}.}
\label{dependence_on_gamma}
\end{figure}

In Fig.\ \ref{dependence_on_gamma} we plot by the dashed (green) line the average induced $y$-component of the angular momentum as a function of the angle of the linear polarization direction of the pulse in the $xz$-plane. We denote this angle by $\gamma$, so that a pulse polarized along the $z$-axis corresponds to $\gamma=0$. It is seen, as expected, that the largest induced $\langle J_y\rangle$ is obtained for a polarization angle close to $45^{\circ}$.

\section{Laser-controlled surface scattering and separation of molecular mixtures}

In this section, we  show that, in principle, one may use  laser-controlled surface scattering for separating molecular beams consisting of several molecular species into individual components.  The scheme seems to be applicable to different types of molecular species, such as isotopes, or nuclear spin isomers. It takes advantage of the fact that rotationally excited and unexcited molecules have different scattering angle distribution after a collision with the surface.

As an example, we  consider a molecular beam composed of a mixture of two nitrogen species. They can be two molecular isotopes, such as $^{14}\text{N}_2$ and $^{15}\text{N}_2$, or two nuclear spin isomers, such as ortho and para isomers of $^{15}\text{N}_2$.
It was shown in the past, both theoretically and experimentally, that two properly delayed ultrashort laser pulses may selectively align a preferred component of such a mixture, while leaving the other one practically unexcited (see \cite{Fleischer06} and \cite{Faucher,Fleischer07}).
\begin{figure}
\centering
\includegraphics[width=0.5\textwidth]{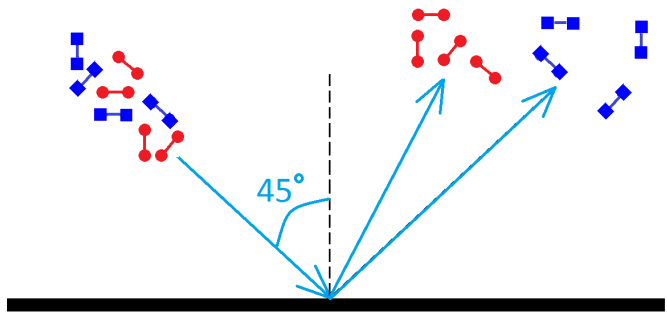}
\caption{(Color online) A mixture of two molecular species is manipulated by two properly delayed ultrashort laser pulses before hitting the surface. The species represented by dumbbells with square ends (blue) are rotationally cold (their rotation is de-excited by the second laser pulse). The species represented by dumbbells with circular ends (red) are rotationally hot (their rotation was further enhanced by the second laser pulse). After hitting the surface, the "blue" species are mostly scattered specularly, and towards angles larger than the specular angle. On the other hand, the "red"  species mostly transfer their rotational energy to translational energy, and are scattered to angles smaller than the specular angle.}
\label{isotope_scheme}
\end{figure}

In order to explain this manipulation, a quantum description of the kicked molecules should be used. After the molecules are kicked by a single laser pulse, they are transiently aligned, and shortly after that become randomly oriented again. However, because the quantum energy levels of the rotor are discrete, and because of the symmetry of a linear molecule, the molecular alignment reappears later due to the phenomenon of quantum revival of the rotational wave packet. Generally, the dynamics of the rotational wave-function  repeats itself after a fixed time period, called the {\it revival time}, which is proportional to the moment of inertia of the molecule.

It was shown in \cite{Fleischer06}, that molecules aligned by the first laser pulse may become even more profoundly  aligned if a second laser pulse is applied to them, which is delayed by an integer multiple of the revival time. However, if the time delay between the two pulses is close to a half-integer multiple of the revival time, the rotational energy given to the molecules by the first pulse is taken away by the second one, and the molecules practically return to the unexcited isotropic state they were in before the first pulse.

Consider now a mixture of two isotopes mentioned above. Because of the mass difference of the isotopes, they have  different revival times. After this mixture is kicked by the first pulse, it is possible to find delay times such that one of the isotopes evolved for an integer number of the revival periods, while the second one completed a half-integer number of its own periods. If the second pulse is applied at one of these moments, then the first isotope will experience enhanced rotational alignment, while the second one will become isotropic and rotationally de-excited.
\begin{figure}
\centering
\includegraphics[width=0.5\textwidth]{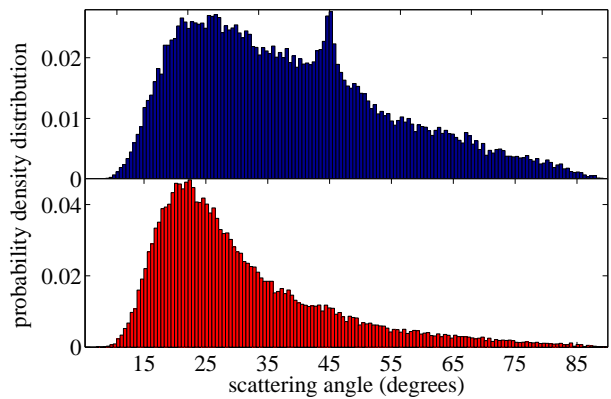}
\caption{ (Color online) The scattering angle distribution of the ``square atom'' (blue) species at the top and the ``circular atom'' (red) species at the bottom, according to the scheme at Fig.\ \ref{isotope_scheme}. The calculation is done using the hard-cube model for nitrogen molecules incident on the surface with translational velocity of $350\,\text{m}/\text{sec}$ and rotational temperature of $1\,\text{K}$. The distribution at the top panel is for the molecules that are not excited by the laser, in order to represent the species that are de-excited by the second laser pulse. The distribution at the bottom panel is for the molecules that are excited by a laser pulse with $P=10$ and $\gamma=0$, to represent the species that are excited twice by a laser pulse with $P=5$. In the upper panel of the figure, only $65\%$ of the molecules are scattered to subspecular angles, while at the lower panel, $84\%$ are scattered to these angles. This means that almost $30\%$ enrichment is achieved for the selectively excited isotope. The surface cube parameters correspond to silver atoms at room temperature ($M=108\,\text{a.u.}$ and $T_{\text{surf}}=300\,\text{K}$).}
\label{isotope_hardwall}
\end{figure}

A similar double pulse approach can be used for selective alignment in a mixture of nuclear spin isomers of $^{15}\text{N}_2$, as explained in \cite{Fleischer07}. In this case, both isomers have the same mass and, therefore, the same revival time. However, kicking the molecules with a second pulse delayed by a time close to a $n+1/4$ (or a $n+3/4$) multiple of the revival time (where $n$ is a positive integer) provides further rotational excitation of  one of the isomers, while de-exciting the other one. The reason for this effect is the entanglement between the molecular rotational and spin degrees of freedom, which is imposed by the Pauli principle (for details, see  \cite{Fleischer07}).

Assume now that such a selectively excited  molecular beam hits a solid surface, as shown schematically in Fig.\ \ref{isotope_scheme}. For a moment, we also assume that the surface is not vibrating, i.e.\ stays frozen at zero temperature. According to Eq.\ (\ref{cons_energy-2}), the rotationally hot (aligned) molecular species, represented by dumbbells with circular ends (red) in Fig.\ \ref{isotope_scheme}, have a high probability of transferring  their rotational energy to translational energy via the collision. As a result, the perpendicular component of molecular velocity  after the collision is larger than the one before the scattering (the parallel velocity component remains the same). In other words, these molecules have a high probability to be scattered to the angles smaller than the specular angle (i.e., to the ``subspecular'' angles). The second species, that are rotationally cold (and isotropically oriented), represented by dumbbells with square ends (blue) in Fig.\ \ref{isotope_scheme}, have a high probability to transfer their translational energy to the rotational energy, and they are mainly scattered to angles larger than the specular angle. As a result of the scattering process, the molecular mixture that arrives to the region of subspecular angles is highly enriched with the ``circular atom'' (red) species. At finite temperature of the surface, the effect is somehow reduced. The reason is that even a non-rotating molecule can acquire translational energy from a vibrating surface atom moving towards it, and scatter to subspecular angles. However, for a strong enough laser pulse, a  sizable effect is expected even for surfaces at room temperature.

To estimate the magnitude of the effect at typical experimental conditions, we consider scattering of a beam consisting of a 1:1 mixture of two nitrogen isotopes discussed above. The beam has initial translational velocity of $350\,\frac{\text{m}}{\text{sec}}$, rotational temperature of $1\,\text{K}$, and it propagates at $45^{\circ}$ with respect to the surface. The surface consists of silver atoms,  and the surface temperature is $300\,\text{K}$. Before hitting the surface, the molecules are excited by a pair of laser pulses of $P=5$ that are polarized along $\mathbf{p}=(0,0,1)$ direction. The timing between the pulses is chosen such that one of the species remains rotationally unexcited, while the other one experiences an efficient kick of $P=10$ from the double pulse. In Fig.\ \ref{isotope_hardwall} we plot the distribution of the scattering angle  for the two molecular isotopes.   The upper panel of the figure corresponds to the  unexcited component of the molecular beam, the lower panel corresponds to the selectively excited isotope.  In both parts of the figure, the distributions are peaked around $25^{\circ}$. This peak comes from the translational energy delivered to the molecules by the thermally oscillating surface atoms.
It is easy to understand the appearance of the maximum around $25^{\circ}$ by using the following simple arguments. Consider a ball with velocity $v$,  colliding with a heavy cube moving towards it with velocity $U$. This ball is reflected from the cube with velocity $v+2U$. Treating the molecule as a ball, and keeping in mind that it hits the cube at angle of $45^{\circ}$, we arrive at a scattering angle of $\arctan{\left(\frac{v/\sqrt{2}}{v/\sqrt{2}+2U}\right)}$. Using a typical value of $\sqrt{k_BT_{\text{surf}}/M}=150\,\frac{\text{m}}{\text{sec}}$ for the velocity $U$, and the incident velocity $v$ of $350\,\frac{\text{m}}{\text{sec}}$ we estimate the scattering angle as $24^{\circ}$, which is consistent with the distribution maxima seen in Fig.\ \ref{isotope_hardwall}.

\begin{figure}
\centering
\includegraphics[width=0.5\textwidth]{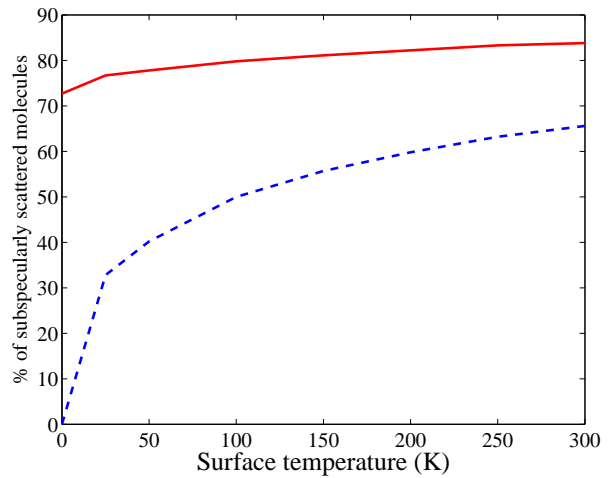}
\caption{(Color online) The percentage of the subspecularly scattered molecules is plotted as a function of the surface temperature for the scattering of the rotationally hot molecules (solid red), and for the rotationally cold molecules (dashed blue). The parameters chosen for the calculation here are the same as in Fig.\ \ref{isotope_hardwall}, except the surface temperature.}
\label{percent_temp}
\end{figure}

Analyzing the distribution functions of Fig.\ \ref{isotope_hardwall}, we find that $84\%$ of the rotationally hot species are scattered to the subspecular angles, while only $65\%$ of the rotationally cold species are scattered to these angles, which means almost $30\%$ enrichment in the selectively excited isotope. This figure is remarkable by itself, however it becomes even more impressive at lower surface temperature. In Fig.\ \ref{percent_temp} we plot the percentage of the subspecularly scattered molecules for the rotationally cold species (dashed blue line), and for the rotationally hot species (solid red line) as a function of the surface temperature. We observe that for any temperature between zero and the room temperature, there are more laser-excited molecules scattered to the subspecular angles, than there are the unexcited ones.  The effect is enhanced dramatically for surface temperature below $50\,\text{K}$, and the ratio between the two isotopes asymptotically tends to the impressive value of 70 at zero surface temperature! This suggests that cooling the solid surface increases the effect to a large extent.

We also explored the dependence of scattering on the laser polarization direction $\gamma$ defined at the end of Section IV. In Fig.\ \ref{dependence_on_gamma} we plot by the solid (blue) line the percentage of the rotationally hot molecules scattered towards subspecular angles as a function of the angle $\gamma$.  We find that the percentage varies slightly with the polarization direction and that the maximal percentage is obtained for pulses polarized at $\gamma=0$.


\section{Conclusions}

We developed a simple classical model, based on the hard cube model, for the description of the molecule-surface scattering process. Using it and the classical model for interaction of molecules with ultrashort laser pulses, we suggested and investigated theoretically two possible schemes for modifying the process of molecule-surface scattering.

In the first scheme, we proposed a  way to exciting unidirectional rotation of molecules. First, the molecules in the molecular beam are aligned by a laser pulse in the direction at some angle with respect to the surface. Then, after the surface scattering, they were shown to rotate preferentially in a direction determined by the laser polarization vector and the surface normal. This anisotropy of the angular momentum of the scattered molecules can be detected, for example, by using the REMPI spectroscopy \cite{Sitz88,Kitano09}.

In the second scheme, we suggested exciting a molecular beam consisting of two molecular species by two properly delayed laser pulses, in order to provide selective rotational excitation of one of the species (similar to \cite{Fleischer06} and \cite{Fleischer07}). We have shown, that these two species are scattered differently from a solid surface, in  a way that  allows for the enrichment of the scattered beam in one of these species for subspecular scattering angles. This result is potentially interesting for the analysis and separation of molecular mixtures of different kinds, including isotopes, and molecular nuclear spin isomers.

\section{Acknowledgments}

Y. K. and I. A. appreciate many fruitful discussions with Eli Pollak, Salvador Miret-Art\'{e}s, Yehiam Prior, Sharly Fleischer and Erez Gershnabel. Financial support for this research from the Israel Science Foundation is gratefully acknowledged. This research is made possible in part by the historic generosity of the Harold Perlman Family.

\appendix*
\section{Three-dimensional scattering of  heteronuclear diatomic molecules from a hard wall}

{\bf Energy conservation.} The conservation of the total (translational and rotational) energy before and after the collision is, similar to Eq.\ (\ref{cons_energy-2}),
\begin{eqnarray}
\frac12(m_1+m_2)V_i^2+\frac12\frac{m_1m_2}{m_1+m_2}r_e^2\omega_i^2 \nonumber \\
=\frac12(m_1+m_2)V_f^2+\frac12\frac{m_1m_2}{m_1+m_2}r_e^2\omega_f^2 \nonumber
\end{eqnarray}
or, using the introduced quantities $\mu=m_1/m_2$ and $v=\frac{\mu}{\mu+1}r_e\omega$:
\begin{equation}
\label{cons_energy_3D}
V_i^2+\frac{1}{\mu} v_i^2=V_f^2+\frac{1}{\mu} v_f^2~.
\end{equation}

{\bf Angular momentum conservation.} The angular momentum conservation in the coordinate system centered at the colliding atom should be expressed now in vector form. The molecule hits the surface when oriented at some $\mathbf{r}_i=\frac{\mu}{\mu+1}r_e(x_i,y_i,z_i)$. The collision is very fast, such that the velocities change, but the orientation remains the same $\mathbf{r}_f=\mathbf{r}_i$ in the course of collision.

We start with the case when the molecule hits the surface with the heavier atom of mass $m_1$, or with $z_i>0$. The angular momentum conservation gives, similar to Eq.\ (\ref{cons_ang_momentum-1}):
\begin{equation}
\label{cons_ang_momentum_3D}
m_2\frac{\mu+1}{\mu}\mathbf{r}_i\times\left(\mathbf{V}_i+\mathbf{v}_i\right)=m_2\frac{\mu+1}{\mu}\mathbf{r}_i\times\left(\mathbf{V}_f+\mathbf{v}_f\right)~,
\end{equation}
or:
\begin{equation}
\label{cons_ang_momentum_3Da}
\mathbf{r}_i\times\left(\mathbf{V}_f-\mathbf{V}_i+\mathbf{v}_f-\mathbf{v}_i\right)=0~.
\end{equation}

It follows from the last equation that the vector in the parentheses should be parallel (or antiparallel) to $\mathbf{r}_i$. The vector $\mathbf{V}_f-\mathbf{V}_i$ is perpendicular to the surface, because we only treat the velocity component perpendicular to the surface (the hard cube is flat and frictionless). The vector $\mathbf{v}_f-\mathbf{v}_i$ should be perpendicular to $\mathbf{r}_i$, because both $\mathbf{v}_i$ and $\mathbf{v}_f$ are perpendicular to $\mathbf{r}_i$. All this leads to the vector diagram in Fig.\ \ref{dumbbell_3D}(b). It follows from the diagram that:
\begin{equation}
\label{cons_ang_momentum_3D_2}
\left|\mathbf{v}_f-\mathbf{v}_i\right|=\left|V_f-V_i\right|\sin{\theta}~,
\end{equation}
an expression similar to Eq.\ (\ref{cons_ang_momentum}), where $V_i$ and $V_f$ denote the center-of-mass velocities with the appropriate sign, and $\sin^2{\theta}=x_i^2+y_i^2$ (in the 3-D case, the angle $\theta$ varies between $0$ and $\pi$, so that $\sin{\theta}$ is always positive). By squaring the last equation, we obtain
\begin{equation}
\label{cons_sq}
v_f^2+v_i^2-2\mathbf{v}_i\cdot\mathbf{v}_f=\left(V_f-V_i\right)^2\sin^2{\theta}~.
\end{equation}
The final velocity $v_f$ can be substituted from Eq.\ (\ref{cons_energy_3D}). The dot product can be found from the following expression:
\begin{equation}
\label{dot_prod}
\left(\mathbf{v}_f-\mathbf{v}_i\right)\cdot\mathbf{v}_i=\mathbf{v}_i\cdot\mathbf{v}_f-v_i^2= \left|\mathbf{v}_f-\mathbf{v}_i\right| v_i \,\vec{e}_{v_i}\cdot\vec{e}_{v_f-v_i}~,
\end{equation}
where $\vec{e}_{v_i}$ and $\vec{e}_{v_f-v_i}$ are the unit vectors in the direction of $\mathbf{v}_i$ and $\mathbf{v}_f-\mathbf{v}_i$, respectively. The unit vector $\vec{e}_{v_f-v_i}$ can be easily found using Fig.\ \ref{dumbbell_3D}(b):
\begin{equation}
\label{unit_vec}
\vec{e}_{v_f-v_i}=-\frac{\vec{e}_z-\left(\vec{e}_z\cdot\mathbf{r}_i\right)\mathbf{r}_i}{\sqrt{1-\left(\vec{e}_z\cdot\mathbf{r}_i\right)^2}}~,
\end{equation}
where $\vec{e}_z$ is the unit vector in the direction perpendicular to the surface (the direction of $\mathbf{V}_f-\mathbf{V}_i$ in Fig.\ \ref{dumbbell_3D}(b)).

Finally, combining Eqs.\ (\ref{cons_energy_3D}), (\ref{cons_ang_momentum_3D_2}), (\ref{cons_sq}) and (\ref{dot_prod}), we obtain for $V_f$, which can be positive or negative:
\begin{equation}
\label{Vf_3d+}
V_f=\frac{\left(\sin^2{\theta}-\mu\right)V_i-2|v_i|\sin{\theta}\left(\vec{e}_{v_i}\cdot\vec{e}_{v_f-v_i}\right)}{\sin^2{\theta}+\mu}~.
\end{equation}
Using Eqs.\ (\ref{cons_ang_momentum_3D_2}), (\ref{unit_vec}) and (\ref{Vf_3d+}) it is easy to find $\mathbf{v}_f$ as well.

Now we analyze the case when the molecule hits the surface with the lighter atom of mass $m_2$, or with $z_i<0$. Here, the energy conservation is the same, but the angular momentum conservation is:
\begin{equation}
\label{cons_ang_momentum_3D-}
-m_1\frac{\mu+1}{\mu}\mathbf{r}_i\times\left(\mathbf{V}_i-\frac{1}{\mu}\mathbf{v}_i\right)=-m_1\frac{\mu+1}{\mu}\mathbf{r}_i\times\left(\mathbf{V}_f-\frac{1}{\mu}\mathbf{v}_f\right)~.
\end{equation}
After a similar derivation, we find:
\begin{equation}
\label{Vf_3d-}
V_f=\frac{\left(\mu\sin^2{\theta}-1\right)V_i-2|v_i|\sin{\theta}\left(\vec{e}_{v_i}\cdot\vec{e}_{v_f-v_i}\right)}{\mu\sin^2{\theta}+1}~,
\end{equation}
and
\begin{equation}
\label{vf_3d-}
\left|\mathbf{v}_f-\mathbf{v}_i\right|=\mu\left|V_f-V_i\right|\sin{\theta}~.
\end{equation}

It can be easily checked that equations (\ref{Vf_3d+}), (\ref{cons_ang_momentum_3D_2}), (\ref{Vf_3d-}) and (\ref{vf_3d-}) reduce to equations (\ref{Vvf_1}) and (\ref{Vvf_2}) for $\mu=1$ and a two-dimensional rotation.

\end{document}